\documentclass[a4paper,20pt]{article}
    \usepackage[T1]{fontenc}
    \usepackage{graphicx}
    \usepackage{psfrag}
   \usepackage{amssymb}
    \usepackage{epsfig}
    \usepackage{amsmath}
    \usepackage{amsfonts}
    \renewcommand{\abstract}{}
\begin{document}
\makeatletter
\renewcommand{\@oddhead}{\textit{YSC'14 Proceedings of Contributed Papers} \hfil \textit{E.P. Kurbatov}}
\renewcommand{\@evenfoot}{\hfil \thepage \hfil}
\renewcommand{\@oddfoot}{\hfil \thepage \hfil}
\fontsize{11}{11} \selectfont

\title{On Star Formation Rate and Turbulent Dissipation in Galactic Models}
\author{\textsl{E.~P. Kurbatov$^{1}$}}
\date{}
\maketitle
\begin{center} {\small $^{1}$Institute of Astronomy, Russian Acad. Sci. \\
kurbatov@inasan.ru}
\end{center}

\begin{abstract}
The models of star formation function and of dissipation of
turbulent energy of interstellar medium are proposed.  In star
formation model the feedback of supernovae is taken into account. It
is shown that hierarchical scenario of galaxy formation with
proposed models of star formation and dissipation is able to explain
the observable star formation pause in the Galaxy.
\end{abstract}

\section*{Introduction}
\indent \indent In models of evolution of galaxies where the direct
gas-dynamical computation is used, it is difficult to reach the
resolution in space less than $10\--100$ pc and resolution in time
less than, say, $10^5$ yr due to computational cost. We can use only
the phenomenological approach to describe the interstellar medium
(ISM) on scale $100$ pc and below.  This means that we have to use
an averaged description of interstellar medium.

In the present paper we consider a star formation model dealing with
turbulent energy of ISM and an averaged approach to take into
account the structure of ISM in model of dissipation of turbulent
energy.

The proposed models are able to explain the observable star
formation pause in our Galaxy corresponding to star ages from
$8\--9$ to $10\--12$ Gyr.

\section*{Star formation model}

\indent \indent The main factor which affects star formation is a
gas density. However, it is clear that star formation can be
suppressed by supernovae explosions by means of turbulent feedback.
Thus, the model star formation law must depend on density and on
turbulent energy of gas.

Let us adopt the star formation law in Kennicutt form
\cite{Kennicutt}:
\begin{equation}
  \psi = c_\ast\,\frac{\rho}{\tau_\mathrm{ff}}  \;,
  \label{eq:schmidt_law}
\end{equation}
where $\rho$ is the gas density, $\tau_\mathrm{ff} \propto
\rho^{-1/2}$ is the free-fall time, $c_\ast$ is the star formation
efficiency (the values are from $0.1$ \cite{Scannapieco} to $1$
\cite{Merlin}).  It is commonly believed that the stars form from a
dense and cool gas.  It is possible to adopt this assumption in a
simple manner.  Just let the star formation efficiency $c_\ast$ be
the fraction of perturbations of density which are unstable by Jeans
criterion, assuming the masses of perturbations distributed in a
power law:
\begin{equation}
  c_\ast \propto m_\mathrm{J}^{1-\beta}  \;,
  \label{eq:sf_efficiency}
\end{equation}
where $m_\mathrm{J} \propto \rho^{-1/2}\,T^{3/2}$ is the Jeans mass
\cite{Jeans} depending on density and ``temperature'' $T$, and
$\beta$ is the exponent of power distribution of mass of
perturbations.  It is necessary to explain the meaning of
temperature in this expression. The temperature value averaged over
all components of ISM $\sim 10^4$K while the temperature of the gas
in star forming regions is tens and units of Kelvin. However, we can
assume the local virial equilibrium between thermal and turbulent
energy of gas, and the power law dependence of turbulent velocity
dispersion on scale, so the value of the averaged turbulent energy
will reflect the value of temperature and turbulent energy on small
scales.

After all substitutions the star formation law will be
\begin{equation}
  \mathrm{SFR} \equiv \psi
  = g\,\rho^{\frac{1}{2}\beta + 1}\,T^{\frac{3}{2}(1 - \beta)}  \;,
  \label{eq:sfr_general}
\end{equation}
or, assuming the Salpeter-like exponent $\beta = 2.35$,
\begin{equation}
  \psi = g\,\frac{\rho^{2.175}}{T^{2.025}} \approx g\,\frac{\rho^2}{T^2}  \;,
  \label{eq:sfr}
\end{equation}
where $g$ is the normalizing constant.  It may be determined using
the star formation law of Tutukov \cite{Tutukov}. According to this
law the star formation process is regulated by ionization and
depends on gas density as
\begin{equation}
  \psi = f \rho^2  \;,\qquad
  f = 2\times10^7\;\text{cm$^3$ g$^{-1}$ s$^{-1}$}  \;.
  \label{eq:sfr_tutukov}
\end{equation}
This law has clear physical base and it works good in the
single-zone model of evolution of galaxies \cite{Wiebe}.  The
average temperature of modern ISM was chosen to be $4\times10^4$ K
(though it's value does not affects significantly on the chemical
abundance history of the Galaxy in single-zone evolution model, see
below), so the normalizing constant becomes $g = 3.2\times10^{16}$
cm$^3$ K$^2$ g$^{-1}$ s$^{-1}$.  The temperature $T$ used in the
last formulae supposed to relate to maximal turbulent velocity
dispersion $\sigma_0^2$ as $\sigma_0^2 = k_\mathrm{B} T / \mu$,
where $k_\mathrm{B}$ is the Boltzmann constant and $\mu$ is the mean
molecular weight.

Some star formation models with turbulent energy account had been
offered earlier \cite{Berczik,Krumholz}, but the model proposed in
this paper has the advantage.  This model based on Jeans instability
creterion so it can be extended for the case when the galaxy
rotation, magnetic field or chemical composition is significant.

\section*{Dissipation model}
\indent \indent In all numerical models of galaxy evolution, the ISM
in a computational cell presented as a solid 'brick' of gas without
any structure.  However, the ISM has complex turbulent behaviors
which are needed to be allowed.  In a simple approach the ISM may be
considered as a set of colliding clouds with different sizes and
masses.  The cloud mass probability distribution function (PDF)
\begin{equation}
  \mathbf{P}\{\mathrm{d}M\} =
    \frac{\mathrm{d}M^{1 - \alpha}}{M_\mathrm{max}^{1 - \alpha}
      - M_\mathrm{min}^{1 - \alpha}}  \;,\qquad
  \alpha \approx 1.5
  \label{eq:cloud_mass_spectrum}
\end{equation}
and density dependence on scale
\begin{equation}
  \rho_l = \rho_0 \left( \frac{l}{l_\mathrm{max}} \right)^{-r}  \;,\qquad
  r \approx 1.1
  \label{eq:cloud_density_spectrum}
\end{equation}
are observable \cite{Smith} and velocity-scale relation had examined
by both observations \cite{Larson} and computations
\cite{Dobbs,Krumholz}:
\begin{equation}
  \sigma_l^2 = \sigma_0^2 \left(\frac{l}{l_\mathrm{max}}\right)^p  \;,\qquad
  p \approx 1  \;.
  \label{eq:velocity_dispersion}
\end{equation}
Here the scale $l$ is in range $(l_\mathrm{min}, l_\mathrm{max})$
and values $\rho_0$ and $\sigma_0$ are averages in an area of size
$l_\mathrm{max}$.

Imagine that dissipation of turbulent energy occurs trough the
collisions of clouds, subsequent contraction and heating by shock
waves and radiation.  It is obvious that the efficiency of
dissipation will determine by relation of collision time to cooling
time \cite{Miniati}.  The value of energy radiated away per unit
time in this process is
\begin{equation}
  Q
  \equiv \left[
    \begin{array}{c}
      \text{dissipation} \\
      \text{rate}
    \end{array}
  \right]
  = \left[
    \begin{array}{c}
      \text{collision} \\
      \text{frequency}
    \end{array}
  \right]  \;
  \sum_{l_\mathrm{min} \leqslant l \leqslant l_\mathrm{max}}
  \left[
    \begin{array}{c}
      \text{energy of clouds} \\
      \text{of size $l$}
    \end{array}
  \right]
  \times \left[
    \begin{array}{c}
      \text{fraction of} \\
      \text{radiated energy}
    \end{array}
  \right]
\end{equation}
The collision rate can be estimated as $\tau_\mathrm{d} =
\sqrt{3/(2\pi G \rho_0)}$ \cite{Firmani, Silk}. The volume density
of turbulent energy of clouds of sizes from $l$ to $l + \mathrm{d}l$
is
\begin{equation}
  \rho_0 \frac{v_l^2 + \sigma_l^2}{2}\,\mathbf{P}\{\mathrm{d}l\} =
  \rho_0 \frac{\sigma_0^2}{2}\,\mathbf{P}\{\mathrm{d}l\}  \;,
  \label{eq:clouds_energy}
\end{equation}
where $v_l^2 = \sigma_0^2 - \sigma_l^2$ is the velocity dispersion
of clouds of size $l$.  $\mathbf{P}\{\mathrm{d}l\}$ is the cloud
size PDF which can be obtained using the simple assumption about
cloud mass and density $M_l = \rho_l l^3$:
\begin{equation}
  \mathbf{P}\{\mathrm{d}l\} =
    \frac{\mathrm{d}l^{1 - \lambda}}%
     {l_\mathrm{max}^{1 - \lambda} - l_\mathrm{min}^{1 - \lambda}}
  \;,\qquad
  \lambda = (\alpha-1) (3-r) + 1 \approx 1.95
  \label{eq:cloud_size_spectrum}
\end{equation}
It is reasonable to describe the fraction of radiated energy for
clouds of size $l$ by Poisson distribution $1 - e^{-q_l}$ where
$q_l$ is the relation of the collision time to cooling time with
cooling function $\Lambda$:
\begin{equation}
  q_l = \tau_{\mathrm{coll}, l}\,
    \frac{\rho_{\mathrm{sh}, l}^2\,\Lambda/\mu^2}{\rho_l \sigma_0^2/2}  \;.
  \label{eq:time_relation}
\end{equation}
Here the cloud density $\rho_{\mathrm{sh}, l}$ behind a shock wave
and the time of collision $\tau_{\mathrm{coll}, l}$ can be
determined for the simple case of head-on collision using the one
dimensional approximation, i.e. Riemann problem \cite{Landau}.  In
this case $\rho_{\mathrm{sh}, l} = \xi_l \rho_l$, where $\xi_l$
depends on $\sigma_0$, $\rho_0$ and $l$, and $\tau_{\mathrm{coll},
l} = l / D_l$, where $D_l$ is a shock velocity which is proportional
to cloud velocity: $D_l = \eta_l v_l$. Finally we can write $Q =
\varepsilon_\mathrm{d} \cfrac{\rho_0 \sigma_0^2}{2 \tau_\mathrm{d}}$
where $\varepsilon_\mathrm{d}$ is the dissipation efficiency which
determines by a complex expression
\begin{equation}
  \varepsilon_\mathrm{d} =
    \frac{1-\lambda}{l_\mathrm{max}^{1-\lambda} - l_\mathrm{min}^{1-\lambda}}
    \int_{l_\mathrm{min}}^{l_\mathrm{max}} \mathrm{d}l\,l^{-\lambda}
      \left( 1 - \exp\left[ -\frac{2 \Lambda l_\mathrm{max}}{\mu^2}\,
    \frac{\rho_0}{\sigma_0^3}\,
    \frac{\xi_l^2}{\eta_l}\,
    \frac{(l/l_\mathrm{max})^{1-r}}{\sqrt{1 - (l/l_\mathrm{max})^p\:}}
    \right] \right)  \;.
  \label{eq:dissipation_efficiency_exact}
\end{equation}
Strong maximum of $\mathbf{P}\{\mathrm{d}l\}$ in $l =
l_\mathrm{min}$, the condition $l_\mathrm{min} \ll l_\mathrm{max}$
and the behaviours of $\xi_l$ and $D_l$ lead to simplified
expression
\begin{equation}
  \varepsilon_\mathrm{d} =
    1 - \exp\left[ -\frac{\Lambda l_\mathrm{max}}{\mu^2}\,
      \frac{\rho_0}{\sigma_0^3} \right]  \;,
  \label{eq:dissipation_efficiency}
\end{equation}
where the single unknown parameter is the maximal scale of
turbulence $l_\mathrm{max}$.  It is set to $5$ pc (this is roughly
the scale of open clusters formation, the turbulence scale cannot be
greater than this).

\section*{Star formation in our Galaxy}
\indent \indent The star formation history in galaxies can have a
strongly non-monotonic nature showing a couple of bursts during a
galaxy life time.  Those bursts can be stimulated by an accretion of
gas from intergalactic medium, collisions or close passing of
galaxies.  The reason to star formation to stop can be the
supernovae explosions \cite{Berman}.  In star formation history of
our Galaxy an epoch exists when the star formation process was
suppressed.  It is seen with distribution of $\mathrm{[Fe/O]}$
\cite{Gratton}, $\mathrm{[Eu/Ba]}\--\mathrm{[Fe/H]}$
\cite{Mashonkina} and $\mathrm{[Mg/Fe]}\--\mathrm{[Fe/H]}$
\cite{Fuhrmann}.  This epoch may be interpreted as a pause between
the end of formation of thick disk and the beginning of formation of
thin disk \cite{Fuhrmann,Mashonkina}. The pause corresponds to star
ages between $8\--9$ and $10\--12$ Gyr.

\begin{figure}[!ht]
  \begin{center}
    \includegraphics[width=16cm]{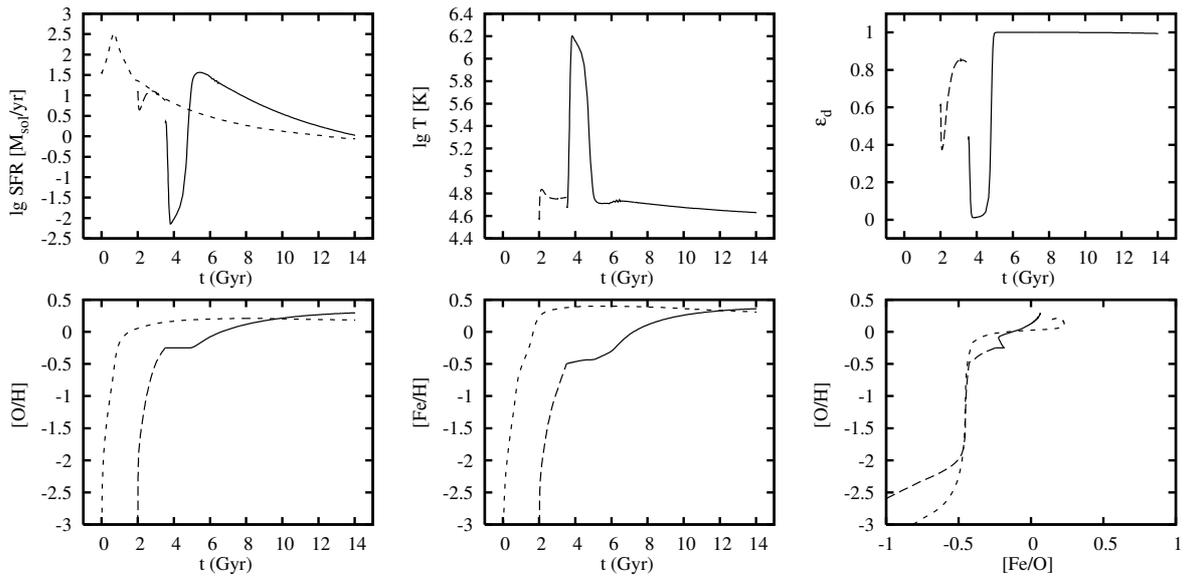}
  \end{center}
  \caption{\footnotesize
    The results of modelling of collision of Galaxy and small satellite.  $SFR$
    is the star formation rate in solar masses per year; $T$ is the
    temperature; $\varepsilon_\mathrm{d}$ is the efficiency of dissipation.
    Dotted line represents the standard model of Firmani \& Tutukov
    \cite{Firmani}, dashed line traces the evolution of a satellite before
    collision and solid line shows the evolution of the collision area in
    Galaxy, after collision.}
  \label{fig:mw}
\end{figure}

The model of star formation and of dissipation proposed above, was
taken to explain this pause in a hierarchical scenario.  These
models was built into the single-zone model of evolution of galaxies
(Firmani \& Tutukov \cite{Firmani}) which was initially implemented
by Wiebe \cite{Wiebe}. The detailed description of modified model
and target setting has given in \cite{Kurbatov}. In modified model
the star formation suppressed by collision of Galaxy and a satellite
of mass $2\times10^{10}$ $M_\odot$ and radius $6.32$ kpc at redshift
$z = 1.5$. The results of modelling are presented on Fig.
\ref{fig:mw}. After collision the temperature increases almost by
two orders causing decrease of dissipation efficiency (see
(\ref{eq:dissipation_efficiency})) with the result that star
formation process ceases due to high temperature (see
(\ref{eq:sfr})) for a $1.5$ Gyr.  The pause in star formation
process is seen as a plateau on the abundance graphs
$\mathrm{[O/H]}$ and $\mathrm{[Fe/H]}$ (solid line).

\begin{figure}[!ht]
  \begin{center}
    \includegraphics[width=8cm]{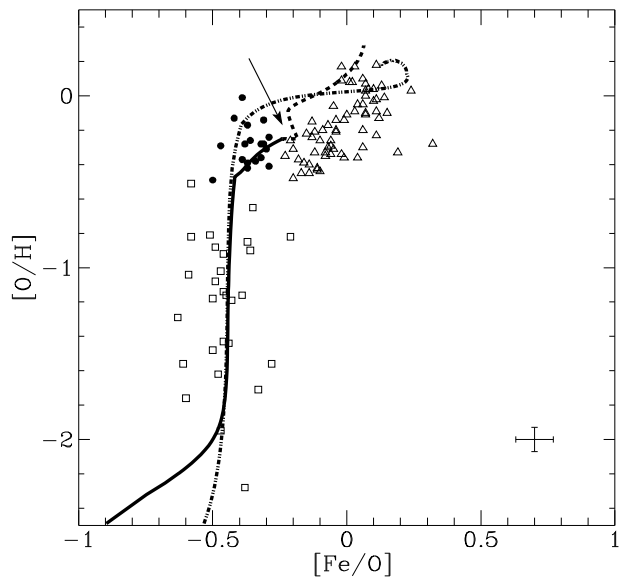}
  \end{center}
  \caption{\footnotesize
    Evolution of chemical abundance of Galaxy in hierarchical scenario.
    Dot-dashed line is the standard model of Firmani \& Tutukov
    \cite{Firmani}, solid line is the satellite and dashed
    line is the collision area in Galaxy.  The arrow marks the position of
    pause in star formation process which is seen as a plateau on the abundance
    graphs on Fig. \ref{fig:mw}.  Symbols are halo stars (open squares), stars
    of the thick disk (filled circles) and thin disk (open triangles).  The
    stars was classified using accurate stellar kinematics
    \cite{Gratton}.  Error bar is at bottom right (there
    are same error bars for all the stars).}
  \label{fig:mw-oh-feo}
\end{figure}

It is interesting to compare the abundance history with distribution
of star abundance in Solar neighborhood.  On Fig.
(\ref{fig:mw-oh-feo}) the $\mathrm{[O/H]}\--\mathrm{[Fe/O]}$
distribution is shown for near metal-poor stars \cite{Gratton}, and
the abundance history in hierarchical scenario.  The arrow marks the
pause of star formation which is close to the bound of populations
of thick disk and thin disk.  The consequence of this pause is
clear: we can see the discontiniuty of the stars age with continuous
abundance history (see the plateau on Fig. \ref{fig:mw}).  The
standard model of Firmani \& Tutukov \cite{Firmani} does not give
such a plateau, giving the right position and shape of the evolution
track, though.

\section*{Conclusions}
\indent \indent The need of taking into account the turbulent energy
in star formation model is obvious.  The large difference of
properties of ISM on different scales is obvious also.  In the
present paper the model of star formation and dissipation of
turbulent energy was offered and implemented to single-zone model of
galaxy evolution. It is shown that the star formation history and
history of chemical abundance in Solar neighborhood can be modeled
in the frame of this model using the scenario of collision of the
Galaxy and the small satellite.

\section*{Acknowledgements}
\indent \indent This work was supported by Russian Foundation for
Basic Research grants 05-02-39005-GFEN\_a and 07-02-00454-a.

\end{document}